\documentstyle[prl,aps]{revtex}
\begin{document}
\draft\onecolumn
\baselineskip=12pt

\title{\bf  Measuring the Cyclotron State of a Trapped Electron}

\author{Stefano Mancini and Paolo Tombesi}

\address{Dipartimento di Matematica e Fisica, 
Universit\`a di Camerino, 
I-62032 Italy\\
Istituto Nazionale di Fisica Nucleare, Sezione di Perugia, Italy\\
and Istituto Nazionale di Fisica della Materia, 
Sezione di Camerino, Italy}

\date{Received: \today}

\maketitle
\widetext

\begin{abstract}
We propose the cyclotron state retrieval of an electron trapped in a 
Penning trap by using different measurement schemes based on 
suitable
modifications of the applied electromagnetic fields and exploiting 
the axial degree of freedom as a probe. A test for matter-antimatter 
symmetry of the quantum state is proposed.
\end{abstract}


\narrowtext

\section{Introduction}
The quantum state of a system is a fundamental concept in quantum 
mechanics, 
because the density matrix describing it contains the complete 
information we can 
obtain about that system. Recently, in quantum 
measurement theory a great interest has been devoted to the 
possibility  
of reconstructing the density matrix by measuring a complete 
set of 
probability distributions over a range of different operator 
representations.  

A tomographic approach to the problem of state retrieval was first 
introduced by J. Bertrand and P. Bertrand \cite{Ber} and some 
measurement schemes proposed to implement quantum state tomography 
are reviewed by Royer \cite{Royer}.
Later on Vogel and Risken \cite{VR} have shown that s-parametrized 
quasiprobability distributions can be obtained from the probability 
distributions of rotated quadrature phases, a technique which allows 
practical implementation within the field of quantum optics 
\cite{Ray}.

The method relies on the possibility of making homodyne 
measurements of 
the field of interest by scanning the phase of the added 
local oscillator 
field, then was named Optical Homodyne Tomography (OHT). 
Recently, 
another method \cite{Wod,WV} was 
introduced based on the direct photon 
counting by scanning both the phase and the amplitude of a 
reference 
field. The latter method, also called Photon Number 
Tomography (PNT) 
\cite{MMT}, has the advantage to avoid sophisticated computer 
processing of the 
recorded data \cite{Wod} and to be applicable when the 
direct access
to the system is inhibited \cite{MMT}.

Although the reconstruction of the phase space distribution was 
already proposed for non-optical systems, i.e. particles 
\cite{Wal,Helon}, 
it is however based on optical measurements performed on the field 
radiated by 
these particles, exploiting the resonance fluorescence phenomenon.
These methods would not be suitable in systems without an internal 
electronic structure
such as the trapped electron (or proton) \cite{Penning}.
The purpose of this work, instead, is to show how to reach the 
characterization of the quantum state of a trapped
"elementary" particle, 
non having an electronic structure,
by using 
tomographic-like measurements without the use of the radiated field. 
In particular, 
we shall consider an electron trapped in a Penning trap 
\cite{Penning}
developing techniques resembling both OHT and PNT, which allow 
to get 
the state of the cyclotron motion by probing its axial degree of 
freedom.
\section{The Model}
We consider the motion of an electron
in a uniform magnetic field $B$ along the positive $z$ axis and a 
static quadrupolar potential. As it is well known \cite{BrownG}, 
the motions of that electron in the trap are well separated 
in energy scale. In what follows we shall consider only the 
cyclotron 
and the axial degrees of freedom, which radiate in the GHz and 
in the  MHz ranges respectively,  neglecting the slow magnetron 
motion in the kHz region. 
To simplify our presentation, we assume the a priori knowledge 
of the
electron's spin \cite{Kells} then, we neglect all the
spin-related  terms in the Hamiltonian that, for an electron of 
rest mass 
$m$ and charge $-|e|$, can be written as the quantum counterpart of 
the classical one
\begin{equation}\label{H1}
\hat H=\frac{1}{2m}\left[{\hat{\bf p}}-\frac{e}{c}
{\hat{\bf A}}\right]^2
+eV_0\,\frac{{\hat x}^2+{\hat y}^2-2{\hat z}^2}{4d^2}\,,
\end{equation}
with ${\hat{\bf A}}=(-{\hat y}B/2,{\hat x}B/2,0)$ and $c$ is the 
speed of light; $d$ characterizes the dimensions of the trap and 
$V_0$ is the potential applied to its electrodes. 

It is convenient to introduce 
the rising and lowering operators for the cyclotron motion,
\begin{eqnarray}\label{rlcyc}
{\hat a}_c&=&\frac{1}{2}\left[\beta({\hat x}-i{\hat y})+
\frac{1}{\beta\hbar}({\hat p}_y+i{\hat p}_x)\right]\,,\\
{\hat a}_c^{\dag}&=&\frac{1}{2}\left[
\beta({\hat x}+i{\hat y})+
\frac{1}{\beta\hbar}({\hat p}_y-i{\hat p}_x)\right]\,,
\end{eqnarray}
with $\beta=(m\omega_c/2\hbar)^{1/2}$ and 
$\omega_c=|e|B/mc$ being the cyclotron angular frequency.
For the axial motion we define
\begin{eqnarray}\label{rlax}
{\hat a}_z&=&\left[\frac{m\omega_z}{2\hbar}\right]^{1/2}
{\hat z}+
i\left[\frac{1}{2m\hbar\omega_z}\right]^{1/2}
{\hat p}_z\,,\\
{\hat a}_z^{\dag}&=&\left[\frac{m\omega_z}{2\hbar}\right]^{1/2}
{\hat z}-
i\left[\frac{1}{2m\hbar\omega_z}\right]^{1/2}{\hat p}_z\,,
\end{eqnarray}
with $\omega_z^2=|e|V_0/md^2$. Thus, by using these new operators, 
the Hamiltonian (\ref{H1}) simply becomes
\begin{equation}\label{H2}
{\hat H}=\hbar\omega_c({\hat a}^{\dag}_c{\hat a}_c+\frac{1}{2})+
\hbar\omega_z({\hat a}^{\dag}_z{\hat a}_z+\frac{1}{2})\,,
\end{equation}
where all terms not containing the rising or loweing operators have 
been omitted. The obtained Hamiltonian (\ref{H2}) is decomposed
into two indipendent parts each forming a complete Hilbert space  
with their own
basis. The state of the electron is thus the inner product of 
the two 
states we shall
call the cyclotron and the axial state, $|\psi\rangle=
|\psi_c\rangle|\psi_a\rangle$.

To go further we would also remark
that the best way of observing the microscopic system from the 
outside 
world is through the measurement of the current due to the induced 
charge on the cap electrodes of the trap, as a consequence of the 
axial motion of the electron along the symmetry axis, 
because there are 
not good detectors in the GHz range to measure 
the cyclotron radiation 
\cite{BrownG}. 
Therefore, since the motions are completely decoupled, only the 
axial 
one is easily detectable \cite{BrownG}. In the following we shall 
consider some 
interaction 
Hamiltonians suitable for an indirect characterization of the 
cyclotron 
motion.
\section{Detection Techniques}
The OHT technique is based on the 
possibility of measuring different quadratures of the field of 
interest; 
let us then define  in our case the generic cyclotron quadrature
\begin{equation}\label{Xc}
{\hat X}_c(\phi)={\hat a}_ce^{-i\phi}+{\hat a}^{\dag}_ce^{i\phi}\,,
\end{equation}
where $\phi$ is the angle in the phase space of the cyclotron 
motion. 
Furthermore, as mentioned above, the axial motion could be 
considered 
as a meter, then in order to couple the meter with the system 
(cyclotron motion), we may consider 
\begin{equation}\label{Hint1}
{\hat H}_{int}=\hbar g\left({\hat a}_ce^{-i\phi+i\omega_ct}
+{\hat a}^{\dag}_ce^{i\phi-i\omega_ct}\right){\hat z}\,,
\end{equation}
where  $g$ is the strenght of interaction. 
This interaction Hamiltonian could be obtained 
by applying the following fields on the trapped particle
\begin{eqnarray}\label{fieldsOHT}
{\hat{\bf A}}&=&\left(\frac{mc}{|e|\beta}g
\hat z\sin(\phi-\omega_ct)
-\frac{B}{2}{\hat y},
\frac{mc}{|e|\beta}g\hat z\cos(\phi-\omega_ct)
+\frac{B}{2}{\hat x},0\right)\,,\nonumber\\
{\hat V}&=&V_0\,\frac{{\hat x}^2+{\hat y}^2-2{\hat z}^2}{4d^2}
-\frac{m}{2e\hbar^2\beta^2}g^2{\hat z}^2\,,
\end{eqnarray}
which differ from the usual ones (Eq. (\ref{H1})) for time 
dependent terms 
added at the pre-existent components.
By considering Eqs. (\ref{H2}), (\ref{Xc}) and (\ref{Hint1}), 
the evolution in the (cyclotron)
interaction picture will be determined by the following Hamiltonian
\begin{equation}\label{HI}
{\hat H}=\hbar\omega_z({\hat a}^{\dag}_z{\hat a}_z+\frac{1}{2})
+\hbar g{\hat X}_c(\phi){\hat z}\,,
\end{equation}
where the cyclotron quadrature phase $\phi$ can be set by the 
experimentalist as shown in Eq.
(\ref{fieldsOHT}).

We now assume that the interaction time is much shorter
than the axial period, i.e. $\tau<< 2\pi/\omega_z$, 
  so that 
the free evolution of the electron in Eq. 
(\ref{HI})
can be neglected. This assumptiom allows to disregard a 
possible detuning of the applied field with respect to
the cyclotron frequency. Indded, in the usual experimental 
set-up the detuning is of the order of few kHz \cite{BrownG}.
 Hence, the effect of the interaction is 
determined by applying a unitary kick
\begin{equation}\label{kick}
\hat K(\tau)=\exp\left[-i g\tau\hat X_c(\phi)\hat z\right] .
\end{equation}

By measuring the 
current due to the induced charge variation on the cap
electrodes of the trap, one gets
the axial momentum \cite{BrownG}, then the 
value of quadrature $\hat X_c(\phi)$ by means of
\begin{eqnarray}\label{pzXc}
\langle\hat p_z(t+\tau)\rangle&=&
\langle{\hat K}^{\dag}(\tau)\hat p_z(t){\hat K}(\tau)\rangle
\nonumber\\
&=&\langle\hat p_z(t)\rangle+
\hbar g\tau\langle\hat X_c(\phi)\rangle\cos(\omega_zt)\,.
\end{eqnarray}
Repeated measurements
allow us to recover the probability distribution ${\cal P}
(X_c,\phi)$ 
for that cyclotron quadrature (marginal distribution). However, 
the measurement process is 
state-destructive, hence the initial state of the electron has to 
be reset prior to each new measurement. After one has performed a 
large set of measurements for each phase angle $\phi$, the 
$s$-parametrized Wigner function can be obtained from the 
probability 
of experimental data ${\cal P}(X_c,\phi)$ through the 
inverse Radon 
transform \cite{VR}
\begin{eqnarray}\label{Radon}
W(\alpha,\alpha^*,s)&=&\int_{-\infty}^{+\infty}
\frac{dr|r|}{4}\int_0^{\pi}\frac{d\phi}{\pi}
\int_{-\infty}^{+\infty}dX_c\,\,{\cal P}(X_c,\phi)\nonumber\\
&\times&\exp\left[sr^2/8+ir\left(x-{\rm Re}(\alpha e^{-i\phi})
\right)\right]\,.
\end{eqnarray}
It is evident from this expression that convergence
problems arise for $s\ge 0$.

In analogy with Ref. \cite{Wod,MMT} we now show how one can 
probe 
the quantum cyclotron phase space by measuring the number of 
cyclotron 
excitations.  
In Ref. \cite{Irene} the quantum non demolition measurement of 
the latter is shown to 
be accessible by only considering the coupling between 
the axial and 
the cyclotron motions induced by the relativistic correction 
to the 
electron's mass; however, in order to get a stronger coupling 
we shall
consider the magnetic bottle configuration, 
which leads to \cite{BrownG}
\begin{equation}\label{Hint2}
{\hat H}_{int}=\hbar\kappa {\hat a}_c^{\dag}{\hat a}_c{\hat z}^2\,,
\end{equation}
where the constant $\kappa$ is related to the strength $b$ of the 
field characterizing the magnetic bottle.
As a consequence of Eq. (\ref{Hint2}), the axial angular frequency 
will be affected by the number of cyclotron excitations, in fact 
it will result
\begin{equation}\label{omz}
{\hat\Omega}_z^2=\omega_z^2+2\frac{\hbar\kappa}{m}
{\hat a}_c^{\dag}{\hat a}_c\,,
\end{equation}
then probing the resonance frequency of the output electric signal 
one can get the number of cyclotron excitations.

The axial motion 
relaxes much faster than the cyclotron one \cite{BrownG}, so,
when the Hamiltonian (\ref{Hint2}) is added to that of Eq. 
(\ref{H1}), 
by considering the axial steady state, and neglecting a very 
small anharmonicity term, we can write
\begin{equation}\label{Hcyc}
{\hat H}_{cyc}=\hbar\left(\omega_c+\kappa\langle{\hat z}^2\rangle
\right){\hat a}_c^{\dag}{\hat a}_c\,,
\end{equation}
with $\langle{\hat z}^2\rangle=k_BT/m\omega_z^2$ the thermal 
equilibrium value of the free axial motion,
from which we may recognize a shift effect also on the cyclotron 
angular frequency due to the coupling. 

Furthermore, to perform a PNT-like,
accordingly to the optical case of Ref. \cite{MMT}, we need of a 
reference field 
which displaces the state one wants to recover or, equivalently, 
which is mixed
to it by means of a beam splitter \cite{Wod,WV} that, however, 
in this case is not applicable. To this end we 
can use a driving field with amplitude $\epsilon$
acting immediately before the measurement process induced by the 
Hamiltonian (\ref{Hint2}), and
given by a Hamiltonian term of the type
\begin{equation}\label{drive}
{\hat H}_{drive}=-i\hbar(\epsilon e^{-i\omega_ct}{\hat a}_c^{\dag}-
\epsilon^*e^{i\omega_ct}{\hat a}_c)\,.
\end{equation}
Hence, in the interaction picture we may write
\begin{equation}\label{rhocyc}
\hat\rho_{cyc}(\tau)={\hat D}^{\dag}(E)\hat\rho_{cyc}(0)
{\hat D}(E)\,,
\end{equation}
with
\begin{equation}\label{D}
\hat D(E)=\exp\left[E{\hat a}_c^{\dag}-
E^*{\hat a}_c\right]\,\quad E=\epsilon\tau\,.
\end{equation}
where $\tau$ represents the driving time interval.
One can again get rid of the detuning with respect to the 
cyclotron frequency provided one chooses the interaction
time $\tau$ much shorter than the axial period.
\section{The Wigner Function}
Once one has "displaced" the initial state, the measurement process 
of the axial frequency
allows to get the number of cyclotron excitations present in the 
state 
$\hat\rho_c(\tau)$, hence the probability  
$P(n_c)=\langle n_c|{\hat D}^{\dag}(E)\hat\rho_{cyc}(0)
{\hat D}(E)|n_c\rangle$ which becomes known $\forall\,n_c$ 
after a large set of measurements.
Then, by referring to the expression of the $s$-parametrized Wigner 
function 
introduced in \cite{Glau}, we may write
\begin{eqnarray}\label{WPNT}
W(E,E^*,s)&=&\frac{2}{1-s}
\sum_{n_c=0}^{\infty}
\left(\frac{s+1}{s-1}\right)^{n_c}\nonumber\\
&\times&\langle n_c|{\hat D}^{\dag}(E)\hat\rho_{cyc}(0){\hat D}(E)
|n_c\rangle\,
\end{eqnarray}
where the quasi-probability distribution corresponds to the state 
$\hat\rho_{cyc}(0)$
and can be entirely obtained by varying the 
complex parameter $E$ (i.e. the reference field).

 This scheme requires the use of the following  fields
\begin{eqnarray}\label{Apnt}
{\hat{\bf A}}=\cases{
\left(2\frac{mc}{\beta |e|}{\rm Im}\{\epsilon e^{-i\omega_ct}\}
-\frac{B}{2}{\hat y},
2\frac{mc}{\beta |e|}{\rm Re}\{\epsilon e^{-i\omega_ct}\}
+\frac{B}{2}{\hat x},
0\right)&$t\le\tau$;\cr
\cr
\left(-\frac{B}{2}{\hat y}-
\frac{b}{2}[{\hat y}{\hat z}^2-{\hat y}^3/3],
\frac{B}{2}{\hat x}+\frac{b}{2}[{\hat x}{\hat z}^2
-{\hat x}^3/3],
0\right)&$t>\tau$;\cr}
\end{eqnarray}
which means to turn on the magnetic bottle as soon as the driving 
field is switched off. The latter may constist 
of an electromagnetic field 
circularly polarized in the $x$-$y$ plane and oscillating at 
$\omega_c$. 
The scalar potential remains the same of Eq. (\ref{H1}), while Eqs. 
(\ref{Hint2}) and (\ref{drive}) 
can be obtained from the above fields by considering $b<<B$ and 
making 
the rotating wave approximation and dipole approximation; in doing 
that one 
also obtain $\kappa=|e|b/2mc$.
We wish to point out that these fields are commonly used in the 
experimental 
set up of the Penning trap, differently from the scheme of 
OHT-like 
where non trivial modifications on the fields are 
requiered.

In conclusion we have shown the possibility of reconstructing the 
cyclotron state 
of a trapped electron using different measurement schemes based 
on suitable 
modifications of the external electromagnetic fields. 
In particular the 
PNT-like could be considered more 
powerfull with 
respect to the OHT-like since does not 
need of any 
filtered back projection process, directly giving the 
desired phase space 
distribution from measured data.
We have confined our treatment to the case of undamped cyclotron
motion which however could be resonable for a perfectly
off resonant situation \cite{GD}.
To simplify the presentation of the measurement schemes we also 
assumed
unity efficiency in the detection process. For nonunity efficiency
it is possible to show \cite{LP,WV} that one never reconstructs the
full Wigner fuction but only a smoothed version of it. 

Moreover, the present model can be applyied to reconstruct 
the quantum
 state of a trapped antiparticle such as the positron 
 (or antiproton) and 
to
 test whether particle and antiparticle have the same quantum state 
under charge conjugation transformation.

Let us give some numerical simulations to emphasize  the discussed 
possibilities of reconstructing quasiprobability distributions. 
We demonstrate the methods assuming that the cyclotron state is an 
odd coherent state state \cite{Manko} (or Schr\"odinger cat state)
\begin{equation}\label{fcat}
|\alpha_-\rangle=N_-(|\alpha\rangle-|-\alpha\rangle)\,,
\end{equation}
with $|\alpha\rangle$ being a coherent state and $N_-$ a 
normalization 
constant. States of this type exhibit quantum interferences 
giving rise 
to negative values and sharp structures in the Wigner function, 
so that 
their reconstruction necessitates particular care.

In Fig. 1 it is shown a smoothed version of the Wigner function 
($s=-0.25$)
reconstructed by using OHT; the negativity of the parameter $s$ 
makes
easier the numerical algorithm for the inversion. 
The full Wigner function ($s=0$), instead, can be readily 
reached by means of PNT, 
Fig. 2, where no convergence problems arise and cumbersome 
numerical
algorithms can be avoided.

Cat states could be created in a Penning trap by considering 
relativistic 
effects (which introduce an anharmonicity), having a macroscopic 
character 
whenever strong excitations are considered, and they can 
be displyed with the aid of the discussed methods \cite{MT}.
From the experimental point of view the difficulty in using the PNT 
is connected with the measurement of the cyclotron excitation 
number  and to distinguish among the Landau level $n_c$ and its
 nearest $n_c \pm 1$. It was shown in \cite{Irene} that with 
the sensibility already reached for the axial resonance frequency, 
this measurement is feasible when the electron's spin is known.

\newpage
FIGURE CAPTIONS

Fig. 1 Simulations of the reconstruction of the 
$s$-parametrized Wigner 
function ($s=-0.25$) for an odd cat state with 
$\alpha=1.5$ by means of 
OHT; 27 phases are scanned with $10^3$ data each.

Fig. 2 Simulations of the reconstruction of the Wigner 
function for an 
odd cat state with $\alpha=1.5$ by means of PNT; $10^3$ 
events are 
sampled for each of the 255 points of the grid.

\end{document}